\documentclass[a4paper,12pt]{article}
\usepackage{epsfig}
\begin{document}
\pagestyle{empty}
\hspace*{12.4cm}MPT-PhT/2000-47\\
\hspace*{13cm}IU-MSTP/44 \\
\hspace*{13cm}hep-lat/0012007 \\
\hspace*{13cm}December, 2000
\begin{center}
 {\Large\bf A Numerical Study of Spectral Flows of Hermitian 
Wilson-Dirac Operator and the Index Theorem in Abelian Gauge 
Theories on Finite Lattices}
\end{center}

\vspace*{1cm}
\def\thefootnote{\fnsymbol{footnote}}
\begin{center}{\sc Takanori Fujiwara}\footnote{Permanent address: 
Department of Mathematical Sciences, Ibaraki University,
Mito 310-8512, Japan} 
\end{center}
\vspace*{0.2cm}
\begin{center}
{\it Max-Planck-Institut f\"ur Physik,
Werner-Heisenberg-Institut, \par
F\"ohringer Ring 6, D-80805 M\"unchen, Germany} 
\end{center}
\vfill
\begin{center}
{\large\sc Abstract}
\end{center}
\noindent
We investigate the index of the Neuberger's Dirac operator in 
abelian gauge theories on finite lattices by numerically analyzing 
the spectrum of the hermitian Wilson-Dirac operator for a continuous 
family of gauge fields connecting different topological sectors. By 
clarifying the characteristic structure of the spectrum leading to 
the index theorem we show that the index coincides to the topological 
charge for a wide class of gauge field configurations. We also argue 
that the index can be found exactly for some special but nontrivial 
configurations in two dimensions by directly analyzing the spectrum. 

\vskip .3cm
\noindent
{\sl PACS:} 02.40.-k, 11.15.Ha

\noindent
{\sl Keywords:} Lattice gauge theory, spectrum, index theorem, 
topological charge

\newpage
\pagestyle{plain}
\setcounter{page}{1}
\noindent
It is generally believed that the nontrivial topological gauge field 
configurations are responsible for nonperturbative aspects such as the 
large $\eta$-$\eta'$ mass splitting in QCD and the fermion number 
violation in the standard model. Lattice gauge theories are the most 
promising approach to such nonperturbative phenomena. However, the 
configuration space of lattice gauge fields is topologically trivial 
since any configuration can be continuously deformed into the trivial 
one. This is not quite disappointing because physically interesting 
gauge fields are smooth and we can impose a kind of smoothness conditions 
that restrict plaquette variables within a small neighborhood of 
unity \cite{lus1}. The space of smooth link variables acquires a 
nontrivial topological structure and it is possible to define 
topological invariants geometrically in terms of lattice gauge 
fields \cite{lus1,phil,PS,FSW}. The index theorem plays a role of a 
touchstone in lattice QCD \cite{KSS,IIY,NV,BDEHT}. 

In the case of Ginsparg-Wilson (GW) Dirac systems \cite{GW,has,neub1} it 
is possible to define exact chiral symmetry \cite{lus2} and the index 
theorem on the lattice relating the index of the GW Dirac operator 
to chiral anomaly is known by Hasenfratz, Laliena and Niedermayer 
(HLN) \cite{HLN,lus2}. Their index theorem is applicable not only 
for any gauge field configurations but also for any GW Dirac operators. 
Since the topological structure of the gauge fields is reflected on 
the very construction of the GW Dirac operator rather implicitly, 
the HLN index theorem tells very few about the relationship between 
the index and the topological invariant of gauge fields for strictly 
finite lattices. Nevertheless we expect that a lattice extension of 
the index theorem relating the index directly with the topological 
invariants of gauge fields can be established for sufficiently smooth 
configurations and physically acceptable GW Dirac operators \cite{HJL}. 
In fact it is possible to relate the axial anomaly (the index density) 
to the topological charge density for abelian theories on the infinite 
lattice quite generally by invoking general principles such as the gauge 
invariance and the locality \cite{lus3}. On strictly finite lattices, 
however, we cannot appeal to the locality. This makes the differential 
geometrical method developed in ref. \cite{lus3} inapplicable. 

In this note we focus our attention to the index of Neuberger's Dirac 
operator \cite{neub1} for compact U(1) theories on finite periodic 
lattices in two and four dimensions and aim at establishing the index 
theorem for smooth background gauge fields.\footnote{The chiral anomalies 
have been established in ref. \cite{KY} in the continuum limit. See also 
ref. \cite{adams}.}  To achieve the goal we 
investigate the spectrum of the related hermitian Wilson-Dirac operators 
numerically for a family of link variables connecting configurations 
with distinct topological charges and clarify the characteristic 
structure of the spectrum leading to the index theorem. Such an analysis 
has already been reported in ref. \cite{NN2} within the framework of the 
overlap formalism. We shall extend the analysis from the point of view of 
Neuberger's Dirac operator to convincing extent. We then show some analytic 
results concerning the spectrum and the index. They are very useful to 
understand the numerical results. 

Let us begin with the definition of Neuberger's Dirac operator on a 
$d=2N$ dimensional euclidean hypercubic regular lattice $L^d$ with periodic 
boundary conditions. For simplicity we choose the lattice spacing $a=1$ 
and take the lattice size $L$ to be a positive integer. We first introduce 
the hermitian Wilson-Dirac operator $H$ by
\begin{eqnarray}
  \label{eq:hWDop}
  H\psi(x)=\gamma_{d+1}\Biggl\{(d-m)\psi(x)-\sum_{\mu=1}^d
  \Biggl(\frac{1-\gamma_\mu}{2}U_\mu(x)\psi(x+\hat\mu)
  +\frac{1+\gamma_\mu}{2}U_\mu(x-\hat\mu)\psi(x-\hat\mu)\Biggr)\Biggr\}~,
\end{eqnarray}
where $m$ is the fermion mass and the Wilson parameter $r=1$ is 
assumed. The $\gamma$-matrices are 
taken to be hermitian and satisfy $\{\gamma_\mu,\gamma_\nu\}
=2\delta_{\mu\nu}$. We use $\gamma_{d+1}=(-i)^\frac{d}{2}\gamma_1
\cdots\gamma_d$. The link variables $U_\mu(x)$ are subject to the 
periodic boundary conditions $U_\mu(x+L\hat\nu)=U_\mu(x)$. We also 
employ the periodic boundary conditions for $\psi(x)$.\footnote{%
We may take anti-periodic boundary condition in the $d$-th lattice 
coordinate. This does not lead to any significant change in the 
numerical analysis.} The Neuberger's Dirac operator $D$ is then given 
by
\begin{eqnarray}
  \label{eq:NDop}
  D=1+\gamma_{d+1}\frac{H}{\sqrt{H^2}}~.
\end{eqnarray}
The $m$ is chosen in the range $0<m<2$ to avoid species doubling. 
We shall often use $m=1$ in the following analysis and denote $H(1)$ 
simply by $H$. Whenever the $m$ dependence is concerned, we write it 
explicitly as $H(m)$. 

{}For the Neuberger's Dirac operator (\ref{eq:NDop}) the index 
theorem of HLN can be stated as 
\begin{eqnarray}
  \label{eq:indth}
  {\rm index} D={\rm Tr}\gamma_{d+1}\Biggl(1-\frac{1}{2}D\Biggr)
  =-\frac{1}{2}{\rm Tr}\frac{H}{\sqrt{H^2}}~.
\end{eqnarray}
This equation relates the index of $D$ to the spectral asymmetry of 
$H$ since the trace on the rhs is nothing but the number of positive 
eigenvalues of $H$ minus that of negative ones \cite{NN1}. It 
enables us to analyze 
the index of the complicated operator $D$ in terms of the much 
simpler operator $H$ that is ultra local on the lattice. 

On finite lattices $D$ is well-defined unless 
$\det H=0$. By excising gauge field configurations leading to $\det H=0$
the space of the link variables
becomes disconnected. We shall refer to the decomposition of the 
configuration space as the analytic. The index theorem (\ref{eq:indth}) 
associates ${\rm index}D$ to each connected component
of the space of link variables. 

The lattice index theorem (\ref{eq:indth}) is valid for any gauge field 
configuration with $\det H\ne0$. However, 
we can restrict ourselves to smooth configurations by imposing
\begin{eqnarray}
  \label{eq:admicond}
  \sup_{x,\mu,\nu}||1-P_{\mu\nu}(x)||\leq \eta ~,
\end{eqnarray}
where $P_{\mu\nu}$
is the standard plaquette variable and $\eta$ is a positive 
constant to be mentioned later. The norm $||A||$ of a matrix $A$ is 
defined by 
\begin{eqnarray}
  \label{eq:matnorm}
  ||A||=\sup_{\varphi,~||\varphi||=1}||A\varphi||~, 
\end{eqnarray}
where $||\varphi||$ is the standard norm for vectors. 
It is well-known that the space of such smooth 
configurations still contains any physically interesting 
gauge field configurations with nontrivial topological 
structure if the size of the lattice is taken to be sufficiently 
large \cite{lus1,lus5}. 

{}For suitable choices of $\eta$ the space of 
link variables becomes disconnected and one can 
define topological invariants geometrically in terms of the link 
variables. We shall refer to the decomposition of the space of 
link variables as the geometric. We describe this in detail for 
compact U(1) theories with which we shall be exclusively concerned. 

In the case of compact U(1) theories (\ref{eq:admicond}) can be 
cast into the simpler form:
\begin{eqnarray}
  \label{eq:aadmicond}
  \sup_{x,\mu,\nu}|F_{\mu\nu}(x)|<\eta' ~.
  \qquad \Biggl(\eta'=2\sin^{-1}\frac{\eta}{2}\Biggr)
\end{eqnarray}
We may choose $0<\eta'<\pi$ for $d=2$ and $0<\eta'<\pi/3$ for $d\geq4$ 
\cite{phil,lus3,lus5,FSW}. The space of link variables is decomposed into a 
finite number of connected 
components characterized by a set of magnetic fluxes $\phi_{\mu\nu}=
2\pi m_{\mu\nu}$ through $\mu\nu$-plane defined by \cite{lus5}
\begin{eqnarray}
  \label{eq:mmunu}
  \phi_{\mu\nu}=\sum_{s,t=0}^{L-1}
  F_{\mu\nu}(x+s\hat\mu+t\hat\nu)~,
\end{eqnarray}
where $m_{\mu\nu}=-m_{\nu\mu}$ is an integer and $F_{\mu\nu}$ is the 
abelian field strength given by 
\begin{eqnarray}
  \label{fmunu}
  F_{\mu\nu}(x)=\frac{1}{i}\ln P_{\mu\nu}(x)~.
  \qquad (|F_{\mu\nu}(x)|<\pi)
\end{eqnarray}
Note that the rhs of (\ref{eq:mmunu}) is independent of $x$ by virtue 
of the Bianchi identity $\partial_{[\lambda}F_{\mu\nu]}=0$. Here 
$\partial_\mu$ is the forward difference operator defined by 
$\partial_\mu f(x)=f(x+\hat\mu)-f(x)$.
The $\phi_{\mu\nu}$ is not only a smooth function 
of the link variables but also is a topological invariant. This implies 
that any two gauge field configurations with distinct sets of magnetic 
fluxes cannot be deformed into each other continuously 
and, hence, belong to different connected components. Conversely, any 
gauge field configurations satisfying (\ref{eq:aadmicond}) can be 
continuously deformed into one another if they have the same set of 
magnetic fluxes \cite{lus5}. This implies that the 
space of gauge fields with a given set of magnetic fluxes that satisfy 
(\ref{eq:aadmicond}) are connected. In particular any gauge field with 
a given set of magnetic fluxes $\phi_{\mu\nu}\equiv2\pi m_{\mu\nu}$ can 
be continuously deformed to a configuration with constant 
field strengths $F_{\mu\nu}(x)=\phi_{\mu\nu}/L^2$. For later convenience 
we denote the space of link variables that satisfy (\ref{eq:aadmicond}) 
and give a set of magnetic fluxes $\phi_{\mu\nu}$ by 
${\cal U}_{\eta,\phi}$.

We can define a topological charge by the lattice analog of Chern 
character \cite{lus3,FSW} as  
\begin{eqnarray}
  \label{eq:topch}
  Q_N&=&\frac{1}{(4\pi)^NN!}\sum_x
  \epsilon_{\mu_1\nu_1\cdots\mu_N\nu_N}F_{\mu_1\nu_1}(x)
  F_{\mu_2\nu_2}(x+\hat\mu_1+\hat\nu_1) \nonumber \\
  &&\hskip 2cm \times\cdots\times 
  F_{\mu_N\nu_N}(x+\hat\mu_1+\hat\nu_1+\cdots
  +\hat\mu_{N-1}+\hat\nu_{N-1})~,
\end{eqnarray}
where $\epsilon_{\mu_1\cdots\mu_d}$ is the Levi-Civita symbol 
in $d$ dimensions. $Q_N$ is a smooth function of the link variables 
within a connected component and takes an integer value given 
by \cite{phil,FSW}
\begin{eqnarray}
  \label{eq:topch2}
  Q_N &=&\frac{1}{2^NN!}\epsilon_{\mu_1\nu_1\cdots\mu_N\nu_N}
  m_{\mu_1\nu_1}m_{\mu_2\nu_2}\cdots m_{\mu_N\nu_N}~.
\end{eqnarray}
A bound on $|Q_N|$ can be easily found from (\ref{eq:topch}) by 
noting (\ref{eq:aadmicond}) as 
\begin{eqnarray}
  \label{eq:Qsb}
  |Q_N|<L^{2N}\frac{\eta'{}^N(2N-1)!!}{(2\pi)^N}~.
\end{eqnarray}

We thus find two different definitions for decomposition 
of the space of link variables, the analytic 
and the geometric. The analytic decomposition is based on the 
requirement $\det H\ne0$ and ${\rm index} D$ is associated to 
each connected component. The geometric decomposition is based on the 
condition (\ref{eq:aadmicond}) and $Q_N$ can be assigned similarly 
to each connected component. In general they are not mutually 
consistent and one can find gauge field configurations that 
satisfy (\ref{eq:aadmicond}) and $\det H=0$ simultaneously. However, 
if we choose $\eta$ within the range
\begin{eqnarray}
  \label{eq:eta}
  0<\eta<\frac{2-\sqrt{2}}{d(d-1)}~,
\end{eqnarray}
$H(=H(1))$ cannot have zero-modes \cite{neub2,HJL} and any connected 
component of 
the space of link variables satisfying (\ref{eq:aadmicond}) is 
contained in some connected component obtained by the analytic 
decomposition. Therefore it is very natural to expect that the 
index of $D$ given by (\ref{eq:indth}) and the topological charge 
(\ref{eq:topch}) coincide with each other. The precise form 
of the index theorem for abelian gauge theories can be stated as : 
\vskip .3cm
\noindent 
{\sl For the link variables satisfying (\ref{eq:aadmicond}) and (\ref{eq:eta}) 
the index of Neuberger's Dirac operator $D$ and the topological charge 
$Q_N$ are related by}
\begin{eqnarray}
  \label{eq:rlid}
  {\rm index} D &=& (-1)^NQ_N~.
\end{eqnarray}
\vskip .3cm
\noindent
To establish (\ref{eq:rlid}) on 
an arbitrary connected component ${\cal U}_{\eta,\phi}$ it suffices to 
show the equality for a particular gauge field configuration with 
constant field strengths $F_{\mu\nu}(x)=\phi_{\mu\nu}/L^2$.

One way to find ${\rm index}D$ for an arbitrary configuration 
is to count the net number of spectral flows of $H(m)$ as $m$ 
varies from $0$ to $1$ \cite{IIY,EHN}. Since $H(m)$ is known to have 
a symmetric spectrum for $m\leq0$ \cite{NN2}, the net number of 
eigenvalues that changes the sign from plus to minus just coincides 
to the index of $D$ at $m=1$ as can be easily seen from (\ref{eq:indth}). 

In what follows we shall take another path since we intend to 
understand the behaviors of ${\rm index}D$ under continuous 
changes of link variables that include topological jumps. 
We consider a one-parameter family of link variables 
$\{U^{(t)}_\mu\}_{0\leq t\leq T}$ connecting two gauge field 
configurations $\{U_\mu^{(0)}\}$ and $\{U_\mu^{(T)}\}$, where both 
topological charge and index are known for $\{U_\mu^{(0)}\}$ but only 
topological charge is known for $\{U_\mu^{(T)}\}$. If $\{U_\mu^{(0)}\}$ 
and $\{U_\mu^{(T)}\}$ belong to two distinct connected components, 
$U_\mu^{(t)}$ goes through configurations violating $\det H\ne0$ 
or (\ref{eq:aadmicond}). We expect jumps in ${\rm index}D$ and $Q_N$ at 
some values of $t$. In the actual analysis we shall choose $U_\mu^{(0)}=1$.
The ${\rm index}D$ for $\{U_\mu^{(T)}\}$ can then be found by counting 
the net number of the eigenvalues of $H$ that change the sign from plus 
to minus as $t$ varies from $0$ to $T$. 

In two dimensions we consider the one-parameter family of link variables 
given by 
\begin{eqnarray}
  \label{eq:2d1pflv}
  U_1^{(t)}(x)=\exp\Biggl[-it\frac{2\pi}{L}x_2\delta_{x_1,L-1}\Biggr]~,
  \qquad
  U_2^{(t)}(x)=\exp\Biggl[it\frac{2\pi}{L^2}x_1\Biggr]~.
  \quad (0\leq x_\mu\leq L-1)
\end{eqnarray}
The field strength is then computed as  
\begin{eqnarray}
  \label{eq:2d1pfs}
  && F_{12}^{(t)}(x)=t\frac{2\pi}{L^2}
  -2\pi(t-n)\delta_{x_1,L-1}\delta_{x_2,L-1} \quad 
  \hbox{for}\quad \frac{(2n-1)L^2}{2(L^2-1)}<t<\frac{(2n+1)L^2}{2(L^2-1)}~,
\end{eqnarray}
where $n$ is an integer with $|n|\leq L^2/2-1$ and is chosen to 
satisfy $|F^{(t)}_{12}(L-1,L-1)|<\pi$. The topological charge $Q_1$ is 
nothing but $m_{12}$ and is given by $Q_1=n$. The link variable 
(\ref{eq:2d1pflv}) connects uniform field strength configurations 
$F^{(n)}_{12}(x)=2\pi n/L^2$ of topological charge $Q_1=n$ 
($n=\pm1,\pm2,\cdots$) continuously to the trivial one 
$F^{(0)}_{12}(x)=0$ for which $H$ has a symmetric spectrum. 
As a function of $t$, $Q_1$ has a discontinuity
at $\displaystyle{t=t^{\rm top}_j\equiv\frac{(2j+1)L^2}{2(L^2-1)}}$ 
($j=0,\pm1,\pm2,\cdots$).

Before going into detail about the numerical results, it is helpful 
to note the following facts: (1) The eigenvalues $\lambda$ of 
$H(m)$ are bounded by $|\lambda|\leq d+|d-m|$ \cite{neub2}.
(2) The eigenvalue spectrum 
of $H(m)$ is $L^2$ periodic in $t$. (3) The eigenvalue spectrum of 
$H(m)$ is symmetric with respect to the point $t=0$, {\it i.e.}, 
$-\lambda$ is an eigenvalue of $H(m)|_{t\rightarrow-t}$ if $\lambda$ 
is an eigenvalue of $H(m)$. From (2) and (3) we see that ${\rm  index}D$ 
is an $L^2$ periodic odd function of $t$ and vanishes at 
$t=0,~\pm  L^2/2$, where the spectrum is symmetric. 

We have analyzed the eigenvalue spectrum of $H$ over the range 
$-L^2/2\leq t\leq L^2/2$ for $2\leq L\leq 15$ by searching for the values 
of $\lambda$ such that ${\rm Re}[\det(H-\lambda)]
{\rm Re}[\det(H-\lambda-\Delta\lambda)]<0$ for small enough 
$\Delta\lambda$.\footnote{In the actual computation 
the approximated values of $\det(H-\lambda)$ have imaginary parts. 
However, the ratios of the imaginary part to the real part are 
very small and the real part is considered to well approximate 
the determinant.} The eigenvalues that are degenerate or approximately 
degenerate with an even multiplicity cannot be detected in this method. 
This, however, causes no problem since for most values of $t$ the 
eigenvalues are not degenerate. 

\begin{figure}[t]
  \begin{center}
    \epsfig{file=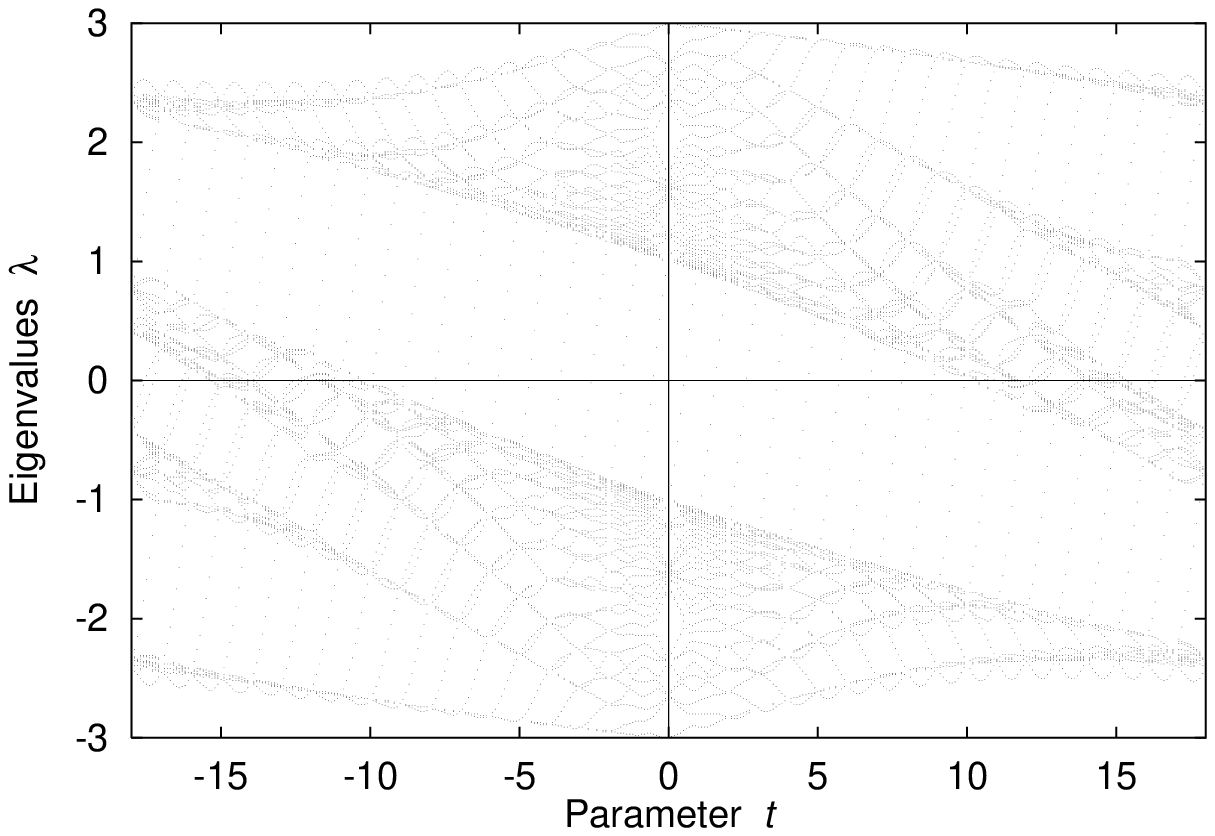,clip=,height=8cm,angle=0}
    \caption{}
    \label{fig:t-2d-l6-wide}
  \end{center}
\end{figure}

In Figure \ref{fig:t-2d-l6-wide} the whole spectrum of $H$ is shown for 
$L=6$.
The characteristic structure of the spectrum is not changed 
with the lattice size $L$. It can be summarized as follows: 
\begin{itemize}
\item Most of the eigenvalues are contained in the upper and the 
lower trapezoid regions symmetrically separated by the parallelogram 
region. 
\item Every time $t$ increases by unity from an integer an eigenvalue 
belonging to the lower trapezoid crosses the parallelogram 
upward and moves to the upper trapezoid. In particular $H$ has 
zero-modes for a sequence of values of $t$ on the interval of the 
horizontal axis cut by the parallelogram. They are almost equally 
spaced. 
\item For integer values of $t$ a kind of clustering of the spectrum 
occurs on the trapezoids and there appear large gaps on the parallelogram. 
In particular, there are $2sL$ eigenvalues with degeneracy $r$ at $t=r$ 
if $L$ is a product of two positive integers $r$ and $s$. 
\end{itemize} 
Let $\kappa(L)$ be the minimum value of $t$ $(>0)$ where an eigenvalue 
changes the sign from plus to minus, then $\kappa(L)$ is larger than 
$L^2/4$. This can be qualitatively understood by noting that 
the smoothed boundary of the parallelogram is convex and goes 
through the points $(0,1)$ and $(L^2/2,-1)$. 
As we shall mention, it is possible to show analytically that for even $L$ 
the eigenvalues of $H$ at $t=L^2/2$ satisfy 
\begin{eqnarray}
  \label{eq:alamr}
  \sqrt{2}-1\leq|\lambda|\leq1~, \qquad 
  \sqrt{5}\leq|\lambda|\leq\sqrt{2}+1~.
\end{eqnarray}
Hence the point $(L^2/2,-1)$ can be regarded as a lower right corner 
point of the upper trapezoid. We see that there is a large gap 
$-\sqrt{5}<\lambda<-1$ on which $H$ has no eigenvalue. It is analogous 
to the gap $-1<\lambda<1$ for the free theory and can be observed at 
integer $t$. 
As for the lower bound on $\kappa(L)$ we also note the exact result 
${\rm index}D=-L^2/4$ for $L$ a multiple of 4. They support the 
aforementioned observations based on the numerical analysis. 

On the interval $-\kappa(L)< t <\kappa(L)$, $2n(L)$ eigenvalues 
flow upward and change the sing at $t^{\rm ind}_j$ $(j=-n(L)+1, 
-n(L)+2,\cdots,n(L)-1,n(L))$ from minus to plus, where 
$t^{\rm ind}_j<t^{\rm ind}_k$ for $j<k$. The ${\rm index}D$ jumps 
by $-1$ at these points and monotonically decreases from the maximum 
$n(L)$ to the minimum $-n(L)$. Since $n(L)\approx \kappa(L)$ for 
large $L$, the maximal index that can be realized by (\ref{eq:NDop}) 
grows faster than $L^2/4$ as $L$ increases. In Table \ref{tab:NL}, 
$n(L)$ is given for $2\leq L\leq15$. We find numerically 
$L^2/n(L)\approx 3.6$ for large $L$. 

\begin{figure}[t]
  \begin{center}
    \epsfig{file=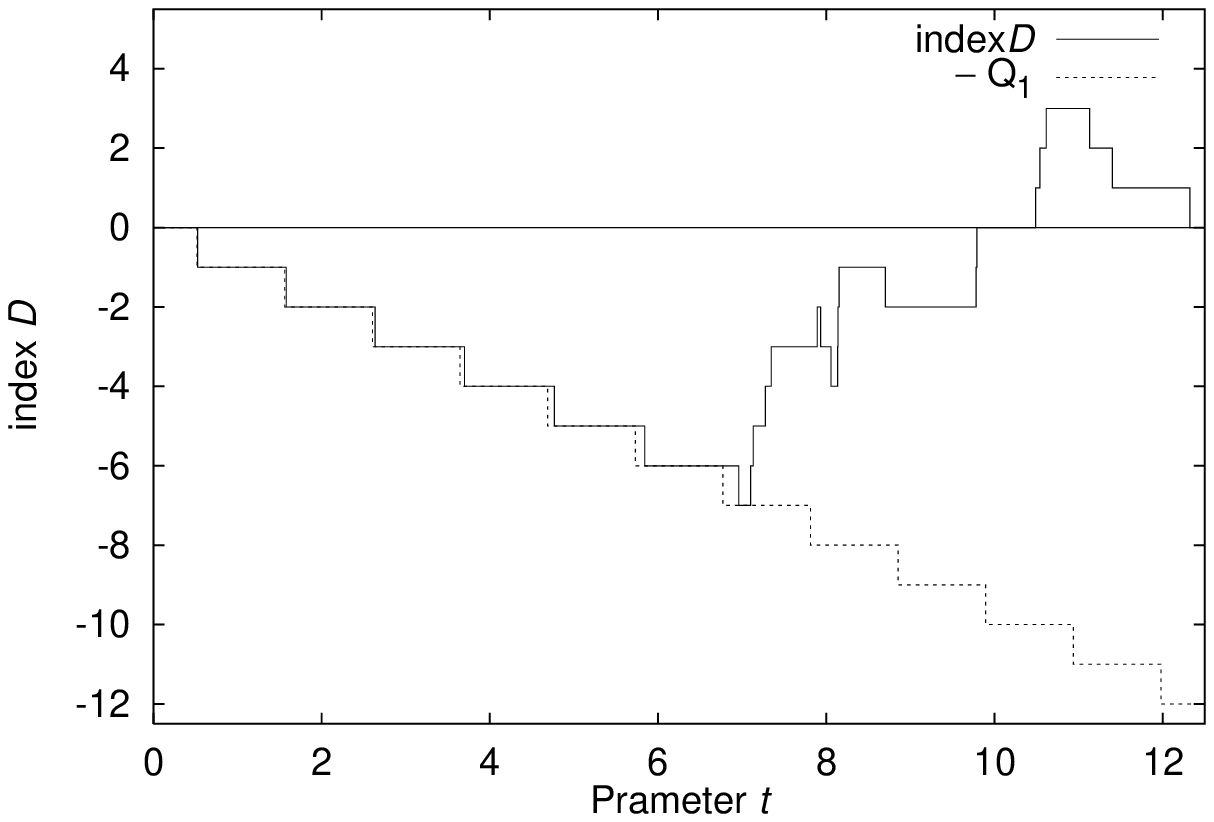,clip=,height=8cm,angle=0}
    \caption{}
    \label{fig:indbarchart}
  \end{center}
\end{figure}

Numerically $t^{\rm ind}_j$ is 
approximately equal to $t^{\rm top}_j$ for $|j|\leq n(L)$ and 
$t^{\rm ind}_j>t^{\rm top}_j$ for $j>0$. In the case of $L=4$, 
for instance, $t^{\rm ind}_j$ ($t^{\rm top}_j$) for $j=1,\cdots,4$ are 
given by $0.5387$ ($0.5333\cdots$), $1.6203$ ($1.6$), $2.7161$ 
($2.666\cdots$), $3.8447$ ($3.7333\cdots$). The difference 
$t^{\rm ind}_j -t^{\rm top}_j$ is monotonically increasing in $j$ for a 
fixed $L$, while it is monotonically decreasing in $L(\geq4)$ for a 
fixed $j(>0)$. We have found that the maximal deviations 
$t^{\rm ind}_{n(L)}-t^{\rm top}_{n(L)}$ are scattered at around $0.2$ 
and $t^{\rm ind}_j$ for $|j|<\hskip -.15cm<n(L)$ approaches to 
$t^{\rm top}_j$ as $L$ becomes large. The numerical values of 
$t^{\rm ind}_j-t^{\rm top}_j$ for fixed $j$ can be well-fitted to a 
curve $\alpha_jL^{-2}$ for large $L$, where the coefficient $\alpha_j$ 
depends on $j$. In the limit $L\rightarrow\infty$, both $t^{\rm ind}_j$ 
and $t^{\rm top}_j$ approach to $\displaystyle{j+\frac{1}{2}}$. 
This is in accord with the observation that on the infinite lattice 
the topological invariants can be investigated quite generally by using 
the geometric concepts \cite{lus3}. 
In Figure \ref{fig:indbarchart} ${\rm index}D$ and $-Q_1$ are plotted 
over the region $0\leq t\leq L^2/2$ for $L=5$. The correspondence 
between the index and the topological charge is very excellent on the 
region $|t|<\kappa(L)$. We thus conclude that ${\rm index}D$ coincides 
to $-Q_1$ for gauge fields sufficiently close to configurations with 
the constant field strength $|F_{12}|\leq2\pi n(L)/L^2$. This is 
consistent with the numerical results given in ref. \cite{chiu}. 

\begin{table}[t]
  \begin{center}
    \begin{tabular}[c]{c|cccccccccccccc}
\hline
      $L$ & 2 & 3 & 4 & 5 & 6 & 7 & 8 & 9 & 10 
      & 11 & 12 & 13 & 14 & 15 \\ 
\hline
      $n(L)$ & 1 & 2 & 4 & 7 & 10 
      & 13 & 18 & 22 & 28 & 34 & 39 & 47 & 55 & 62
\\ 
      $n_\eta(L)$ & 0 & 0 & 0 & 1 & 1 & 2 & 2 & 3 
      & 4 & 5 & 6 & 7 & 9 & 10
\\ \hline
    \end{tabular}
    \caption{}
    \label{tab:NL}
  \end{center}
\end{table}

The jumps of ${\rm index}D$ and those of $-Q_1$, however, do not 
exactly synchronize and the equality between ${\rm index}D$ and $-Q_1$ 
is violated near the discontinuities, where the field strength 
(\ref{eq:2d1pfs}) at the corner point $x_1=x_2=L-1$ takes a large value 
$\approx\pm\pi$. The finite deviations between $t^{\rm ind}_j$ and 
$t^{\rm top}_j$ imply that the analytic decomposition of the space of 
link variables into connected components by imposing $\det H\ne0$ and 
the geometric one by imposing (\ref{eq:aadmicond}) are not consistent 
with each other for a general choice of $\eta$. If $\eta$ is chosen to 
satisfy $0<\eta<1-1/\sqrt{2}$ as given by (\ref{eq:eta}), the 
condition (\ref{eq:aadmicond}) excludes uniformly the regions where the 
jumps both in the index and in the topological charge occur. We thus 
find that ${\rm index}D$ coincides with $-Q_1$ for any gauge fields 
contained in ${\cal U}_{\eta,\phi}$ and the index theorem (\ref{eq:rlid}) 
holds true in two dimensions. On the configuration space 
${\cal U}_{\eta,\phi}$ the maximal index is given by $[\eta' L^2/2\pi]$. 
In Table \ref{tab:NL} we give bounds $n_\eta(L)$ on the index for 
$2\leq L\leq15$ for comparison with $n(L)$. 

The spectral flow crossing $\lambda=0$ is complicated 
on the interval $\kappa(L)\leq t\leq L^2/2$. The eigenvalues flow both 
downward and upward. The downward flows are dense and concentrated in 
narrow regions, where the index increases rapidly. Then the index 
decreases slowly due to the sparse upward flows until the next dense 
downward flows approach to the horizontal axis. The downward flows 
and the upward flows are repeated alternately depending on the size 
of the lattice. Finally we have ${\rm index}D=0$ at $t=L^2/2$. 
Similar thing also happens on the interval $-L^2/2\leq t\leq-\kappa(L)$. 
We should consider, however, such peculiar behaviors on the outer regions 
$\kappa(L)\leq|t|\leq L^2/2$ as a kind of lattice artifacts
and the corresponding field configurations as unphysical. 
In other words the field configurations should be restricted not 
to involve such lattice artifacts in order to keep proper connection 
with continuum theories. 

\begin{figure}[t]
  \begin{center}
    \epsfig{file=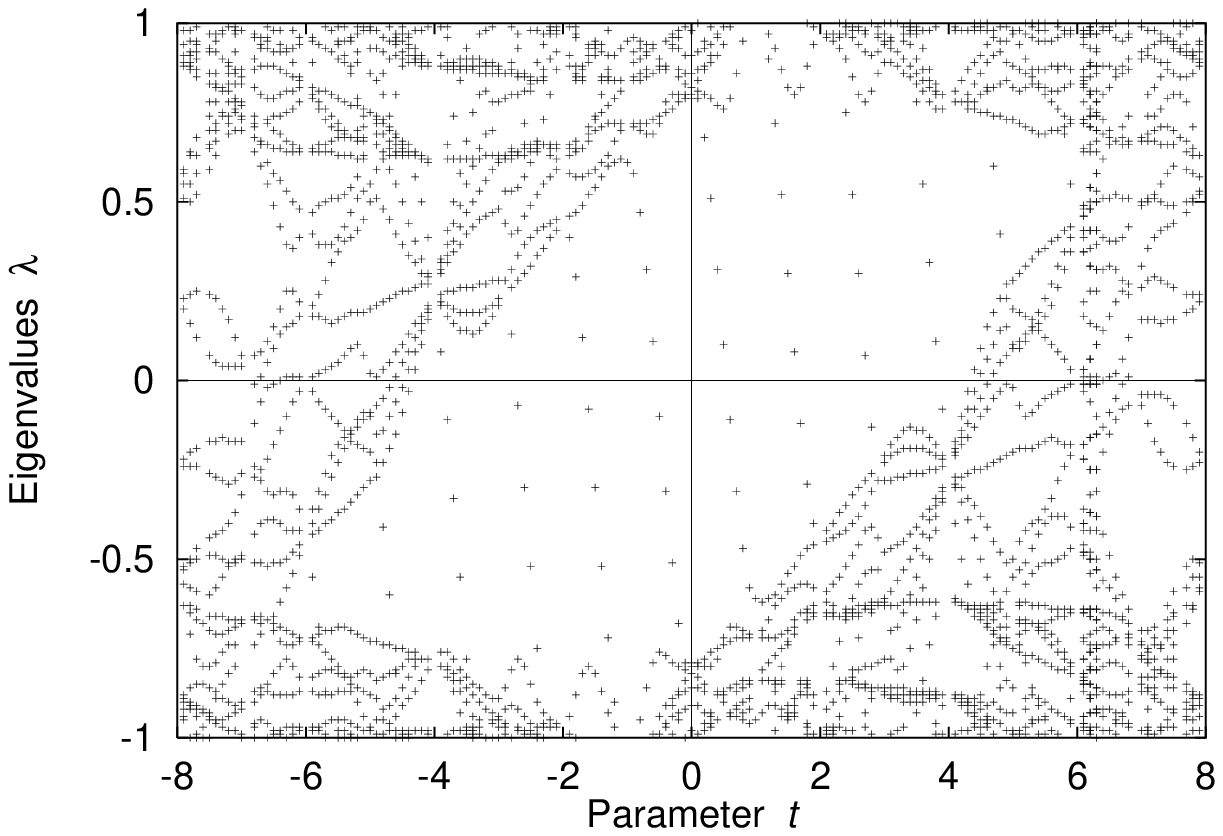,clip=,height=8cm,angle=0}
    \caption{}
    \label{fig:t-4d-l4}
  \end{center}
\end{figure}

We have carried out a similar analysis in four dimensions 
by taking the link variables (\ref{eq:2d1pflv}) and 
\begin{eqnarray}
  \label{eq:34lv}
  U^{(s)}_3(x)=\exp\Biggl[-is\frac{2\pi}{L}x_4\delta_{x_3,L-1}\Biggr]~, 
  \qquad U^{(s)}_4(x)=\exp\Biggl[is\frac{2\pi}{L^2}x_3\Biggr]~,
\end{eqnarray}
where $s$ is an integer ($|s|\leq L^2/2$) corresponding to $m_{34}=s$. 
Only the link variables along the first two lattice coordinates are 
changed. Since the spectrum of $H$ is symmetric at $t=0$, we can 
find ${\rm index}D$ as in two dimensions. The characteristic features 
of the spectral flows observed in two dimensions can be seen also in 
four dimensions. In Figure \ref{fig:t-4d-l4} we indicate the spectrum 
of $H$ for $L=4$, $|\lambda|\leq1$, $s=1$ and $|t|\leq8$. Most of 
the eigenvalues are distributed outside the parallelogram region and 
an eigenvalue crosses the parallelogram downward as $t$ increases by 
unity from an integer. When it crosses the horizontal axis, 
${\rm index}D$ increases by one unit. Note that the direction of the 
eigenvalue flows on the parallelogram region is consistent with the 
factor $(-1)^N$ on the rhs of (\ref{eq:rlid}). The characteristic 
feature of the spectrum is not changed for $s>1$. The parallelogram 
region becomes smaller as $s$ increases. In general $s$ adjacent 
eigenvalues flow downward and ${\rm index}D$ increases by $s$ when 
they cross the $t$-axis. This is consistent with (\ref{eq:topch2}). 
We found that the index and the topological charge coincide with each 
other for the values of $(s,t)$ ($0<s\leq t$) given in Table \ref{tab:st}. 
Though our analysis is restricted to rather small lattice sizes 
$2\leq L\leq4$, we anticipate that the parallelogram region expands 
rapidly enough as $L$ increase and the condition (\ref{eq:aadmicond}) 
with $0<\eta<(2-\sqrt{2})/12$ is sufficient to ensure the lattice 
index theorem (\ref{eq:rlid}) in four dimensions.\footnote{For gauge 
fields satisfying (\ref{eq:aadmicond}) and (\ref{eq:eta}) nonvanishing 
topological charges can be realized only for $L\geq 9$.}

\begin{table}[t]
  \begin{center}
    \begin{tabular}[c]{c|ccc}
\hline
      $L$ & 2 & 3 & 4 \\ 
\hline
      $(s,t)$ & none & (1,1) & (1,1), (1,2), (1,3), (1,4), (2,2), (2,3) \\ 
\hline
    \end{tabular}
    \caption{}
    \label{tab:st}
  \end{center}
\end{table}

Coming back to two dimensions, we give some analytic 
results on the spectrum of $H$ and 
${\rm index}D$ at $t=r, L^2/r$, where $r$ is an arbitrary integer 
factor of $L$.
They are helpful to understand our numerical results. 

The eigenvalues $\lambda$ of $H$ at $t=L^2/r$ with $L$ a multiple of 
$r$ are parametrized by $p$ and $q$ 
\begin{eqnarray}
  \label{eq:pqr}
  p=\frac{2\pi k}{L}~,\quad q=\frac{2 r\pi l}{L}~,
  \qquad \Biggl(\Biggl[-\frac{L}{2}\Biggr]< k\leq
  \Biggl[-\frac{L}{2}\Biggr]~, \quad 
  0\leq l< \Biggl[\frac{L}{r}\Biggr]\Biggr)
\end{eqnarray}
and satisfy the secular equation \cite{fuji}
\begin{eqnarray}
  \label{eq:ceq}
  \det\pmatrix{B(p,q)-\lambda& C(p,q) \cr
    C(p,q)^\dagger & -B(p,q)-\lambda}=0~,
\end{eqnarray}
where $B(p,q)$ and $C(p,q)$ are $r\times r$ matrices defined by
\begin{eqnarray}
  \label{eq:BCme}
  && (B(p,q))_{kl}=-\frac{1}{2}\delta^{(q)}_{k,l+1}
  +\Biggl[1-\cos\Biggl(p+\frac{2\pi k}{r}\Biggr)\Biggr]\delta^{(q)}_{k,l}
  -\frac{1}{2}\delta^{(q)}_{k+1,l} \nonumber \\
  && (C(p,q))_{kl}=-\frac{1}{2}\delta^{(q)}_{k,l+1}
  +\sin\Biggl(p+\frac{2\pi k}{r}\Biggr)\delta^{(q)}_{k,l}
  +\frac{1}{2}\delta^{(q)}_{k+1,l}~. 
  \quad (0\leq k,l\leq r-1)
\end{eqnarray}
The $\delta^{(q)}_{k,l}$ is the Kronecker's $\delta$-symbol for 
$0\leq k,l\leq r-1$ and satisfies the twisted boundary conditions 
$\delta^{(q)}_{k,r}={\rm e}^{-iq}\delta_{k,0}$ and $\delta^{(q)}_{r,k}
={\rm e}^{iq}\delta_{0,k}$. 
It can be shown that (\ref{eq:ceq}) takes the following form \cite{fuji}
\begin{eqnarray}
  \label{eq:ceqgf}
  f_r(\lambda)=\frac{(-1)^{r-1}}{2^{r-4}}\sin^2\frac{rp}{2}\sin^2\frac{q}{2}~,
\end{eqnarray}
where $f_r=\lambda^{2r}+\cdots$ is a polynomial of degree $2r$ and is 
independent of $p$ and $q$. For $1\leq r\leq6$ it is given by 
\begin{eqnarray}
  \label{eq:frs}
  f_1&=&\lambda^2-1~, \nonumber \\
  f_2&=&\lambda^4-6\lambda^2+1~,\nonumber \\
  f_3&=&\lambda^6-9\lambda^4
  +\frac{69}{4}\lambda^2-3\sqrt{3}\lambda-\frac{1}{4}~,\nonumber \\
  f_4&=&\lambda^8-12\lambda^6+42\lambda^4-8\lambda^3-44\lambda^2
  +24\lambda-3~, \nonumber \\
  f_5&=&\lambda^{10}-15\lambda^8+\frac{595+5\sqrt{5}}{8}\lambda^6
  -\frac{5\sqrt{10+2\sqrt{5}}}{2}\lambda^5\nonumber \\
  &&-\frac{1095+65\sqrt{5}}{8}\lambda^4 
  +15\sqrt{10+2\sqrt{5}}\lambda^3+\frac{1115+270\sqrt{5}}{16}\lambda^2
  \nonumber \\
  && -\frac{(85+35\sqrt{5})\sqrt{10+2\sqrt{5}}}{8}\lambda 
  +\frac{109+50\sqrt{5}}{16}~, \nonumber \\
  f_6&=&\lambda^{12}-18\lambda^{10}+\frac{237}{2}\lambda^8
  -6\sqrt{3}\lambda^7-360\lambda^6+54\sqrt{3}\lambda^5
  \nonumber \\
  &&+\frac{8145}{16}\lambda^4-\frac{297\sqrt{3}}{2}\lambda^3
  -\frac{2007}{8}\lambda^2+\frac{255\sqrt{3}}{2}\lambda-\frac{719}{16} ~.
\end{eqnarray}
The case of $r=1$ corresponds to the free spectrum since the link variables 
become trivial.

The eigenvalues must satisfy the inequality
\begin{eqnarray}
  \label{eq:evineq}
  0\leq (-1)^{r-1}2^{r-4}f_r(\lambda)\leq 1~.
\end{eqnarray}
This gives rise to $2r$ allowed intervals 
$\lambda^{(0)}_s\leq \lambda\leq\lambda^{(1)}_s$ ($s=1,\cdots,2r$), 
where $\lambda^{(0)}_s$ and $\lambda^{(1)}_s$ are the $s$-th largest 
roots of either $f_r(\lambda)=0$ or $f_r(\lambda)=(-1)^{r-1}/2^{r-4}$. 
In general the allowed intervals become shorter and shorter as $r$ 
increases. This is the reason why eigenvalues form clusters around the 
point where $t=L^2/r$ takes an integer. In particular we find 
(\ref{eq:alamr}) from (\ref{eq:evineq}) for $r=2$. Similar mechanism 
of clustering is also at work for other integer points $t$ $(\ne 1)$ 
as is seen in Figure \ref{fig:t-2d-l6-wide}. 

If $f_r$ is known, it is possible to find ${\rm index}D$ at $t=L^2/r$. 
As an example, we consider the case that $L$ is a multiple of 4 and take 
$r=4$. For any $p,q$ the eight roots of (\ref{eq:ceqgf}) are separately 
located in the following narrow intervals
\begin{eqnarray}
  \label{eq:lnr}
  &&-2.4142\cdots\leq\lambda\leq-2.4087\cdots~, \qquad
  -1.9429\cdots\leq\lambda\leq-1.9032\cdots~, \nonumber \\
  &&-1.7320\cdots\leq\lambda\leq-1.6914\cdots~, \qquad
  0.1029\cdots\leq\lambda\leq0.1939\cdots~, \nonumber \\
  &&0.4142\cdots\leq\lambda\leq0.5609\cdots~, \qquad
  \hskip .65cm 0.9260\cdots\leq\lambda\leq1~, \nonumber \\
  &&1.7320\cdots\leq\lambda\leq1.7448\cdots~,\qquad
  \hskip .65cm 2.7082\cdots\leq\lambda\leq2.7092\cdots~. 
\end{eqnarray}
It has three negative roots and five positive roots for each $p,q$, 
giving the exact index $-1\times L\times L/4=-L^2/4$. Note also that 
there is a large gap between the largest negative root 
and the smallest positive root. In general (\ref{eq:ceqgf}) has $r-1$ 
negative roots and $r+1$ positive roots for $r\geq4$ and, hence, 
${\rm index}D=-L^2/r$ at $t=L^2/r$. On the other hand it is vanishing 
for $1\leq r\leq3$ since the number of positive roots and that 
of negative ones are always equal. 

These results can be extended to integers $t=r=L/s$ $(\leq L)$ by noting 
the relation \cite{fuji}
\begin{eqnarray}
  \label{eq:h-lfr}
  \det(H-\lambda)|_{t=r}=(f_{sL}(\lambda))^r~,
\end{eqnarray}
where $r$ and $s$ are positive integers satisfying $L=rs$. This gives 
at $t=r$ the index $-r$ for $sL\geq4$ and 0 for $sL\leq3$. 
These are completely consistent with the numerical results. 

We have confirmed that the equality (\ref{eq:rlid}) between the index 
and the topological charge holds true for a wide class of gauge fields 
on the finite periodic lattice in two and four dimensions. The condition 
(\ref{eq:aadmicond}) with (\ref{eq:eta}) excludes uniformly the 
configurations for which the discrepancy between the index and the 
topological charge appears and ensures the index theorem (\ref{eq:rlid}). 
We should give a remark on the possibility of extending our analysis to 
nonabelian theories with nontrivial topological charges. Link variables 
with nontrivial topological charges can be easily found within the 
commuting subgroup. The analysis can be reduced to that of a multi-U(1) 
cases and our results can be applied straightforwardly. The index theorem 
(\ref{eq:rlid}) also has an implication on chiral 
gauge theories on finite periodic lattices. We will argue them elsewhere. 

\vskip .3cm
I would like to thank Peter Weisz for giving me the valuable suggestions 
to improve my primitive numerical results. I am also grateful to Dieter 
Maison and Theory Group for the kind hospitality during my stay at 
Max-Planck-Institute. 

\vskip .3cm


\begin{thebibliography}{99}
\bibitem{lus1} M. L\"uscher, Commun. Math. Phys. {\bf 85} (1982) 39.
\bibitem{phil} A. Phillips, Ann. Phys. {\bf 161} (1985) 399.
\bibitem{PS}
M. G\"ockeler, A.S. Kronfeld, M.L. Laursen, G. Schierholz and 
U.-J. Wiese, Nucl. Phys. {\bf B292} (1987) 349.
\bibitem{FSW} T. Fujiwara, H. Suzuki and K. Wu, IU-MSTP/39, hep-th/0001029.
\bibitem{KSS} F. Karsch, E. Seiler ans I.O. Stamatescu, Nucl. Phys. 
{\bf B271} (1986) 349;
\newline
J. Smit and J.C. Vink, Nucl. Phys. {\bf B286} (1987) 485.
\bibitem{IIY} S. Itoh, Y. Iwasaki and T. Yoshi\'e, Phys. Rev. {\bf D36}
(1987) 527.
\bibitem{NV} R. Narayanan and P. Vranas, Nucl. Phys. {\bf B506} (1998) 
373.
\bibitem{BDEHT} W. Bardeen, A. Duncan, E. Eichten, G. Hockney and 
H. Thacker, Phys. Rev. {\bf D57} (1998) 1633;
\newline
C. Gattringer and I. Hip, Nucl. Phys. {\bf B536} (1998) 363;
\newline
P. Hern\'andez, Nucl. Phys. {\bf B536} (1998) 345.
\bibitem{GW} P.H. Ginsparg and K.G. Wilson, Phys. Rev. {\bf D25} 
(1982) 2649.
\bibitem{has} P. Hasenfratz, Nucl. Phys. (Proc. Suppl.) {\bf 63} 
(1998) 53; 
Nucl. Phys. {\bf B525} (1998) 401.
\bibitem{neub1} H. Neuberger, Phys. Lett. {\bf B417} (1998) 141; 
{\bf B427} (1998) 353.
\bibitem{lus2} M. L\"uscher, Phys. Lett. {\bf B428} (1998) 342.
\bibitem{HLN} P. Hasenfratz, V. Laliena and F. Niedermayer, 
Phys. Lett. {\bf B427} (1998) 125.
\bibitem{HJL} P. Hern\'andez, K. Jansen and M. L\"uscher, 
Nucl. Phys. {\bf B552} (1999) 363.
\bibitem{lus3} M. L\"uscher, Nucl. Phys. {\bf B538} (1999) 515;
\newline
T. Fujiwara, H. Suzuki and K. Wu, Phys. Lett. 
{\bf B463} (1999) 63; 
Nucl. Phys. {\bf B569} (2000) 643; IU-MSTP/37, hep-lat/9910030. 
\bibitem{KY} Y. Kikukawa and A. Yamada, Phys. Lett. {\bf B448} 
(1999) 265;
\newline
D.H. Adams, hep-lat/9812003;
\newline
H. Suzuki, Prog. Theor. Phys. {\bf 102} (1999) 141. 
\bibitem{adams} D. H. Adams, Chin. J. Phys. {\bf 38} (2000) 633. 
\bibitem{NN2} R. Narayanan and H. Neuberger, Phys. Rev. Lett. 
{\bf 71} (1993) 3251; Nucl. Phys. {\bf B443} (1995) 305.
\bibitem{NN1} R. Narayanan and H. Neuberger, Nucl. Phys. {\bf B412} 
(1994) 574.
\bibitem{lus5} M. L\"uscher, Nucl. Phys. {\bf B549} (1999) 295.
\bibitem{neub2} H. Neuberger, Phys. Rev. {\bf D61} (2000) 085015.
\bibitem{EHN} R. Edwards, U. Heller and R. Narayanan, Nucl. Phys. 
{\bf B522} (1998) 285.
\bibitem{chiu} T.W. Chiu, hep-lat/9911010.
\bibitem{fuji} T. Fujiwara, in preparation.
\end{thebibliography}
\end{document}